\documentclass[conference]{IEEEtran}
\usepackage{amsmath}
\usepackage{latexsym}
\usepackage{graphicx}
\usepackage{bbding}
\usepackage{indentfirst}
\IEEEoverridecommandlockouts
\begin{document}

\title{Exploiting Large-Scale MIMO Techniques for Physical Layer Security with Imperfect Channel State Information}
\author{\authorblockN{Xiaoming~Chen$^{\dagger,\ddagger}$,
Chau~Yuen$^{\ast}$, Zhaoyang Zhang$^{\star}$
\\$^{\dagger}$ College of Electronic and Information Engineering, Nanjing
University of Aeronautics and Astronautics, China.
\\$^{\ddagger}$ National Mobile Communications Research Laboratory, Southeast
University, China.
\\$^{\ast}$ Singapore University of Technology and Design, Singapore.
\\$^{\star}$ Department of Information Science and Electronic Engineering, Zhejiang University, China.
\\Email: chenxiaoming@nuaa.edu.cn, yuenchau@sutd.edu.sg,
ning\_ming@zju.edu.cn
}} \maketitle

\begin{abstract}
In this paper, we study the problem of physical layer security in
large-scale multiple input multiple output (LS-MIMO) systems. The
large number of antenna elements in LS-MIMO system is exploited to
enhance transmission security and improve system performance,
especially when the eavesdropper is closer to the information
source and has more antennas than the legitimate user. However, in
practical systems, the problem becomes challenging because the
eavesdropper channel state information (CSI) is usually
unavailable without cooperation and the legitimate CSI may be
imperfect due to channel estimation error. In this paper, we first
analyze the performance of physical layer security without
eavesdropper CSI and with imperfect legitimate CSI, and then
propose an energy-efficient power allocation scheme to meet the
demand for wireless security and quality of service (QoS)
simultaneously. Finally, numerical results validate the
effectiveness of the proposed scheme.
\end{abstract}

\section{Introduction}
The broadcast nature of wireless channels enables the freedom to
communicate at anywhere, but also leads to information leakage.
Traditionally, information security is realized by using
cryptography technology. In fact, secrecy transmission can be
realized through physical layer security by exploiting the channel
fading and noise, which avoids the utilization of cryptography
methods \cite{Wyner} \cite{PLS1}. In brief, as long as the
eavesdropper channel is degraded, it is likely to provide the secure
transmission based on physical layer security.

The performance of physical layer security is usually measured by
secrecy rate, defined as the difference between the legitimate
channel capacity and the eavesdropper channel capacity
\cite{Multiantenna1}. It is generally known that multi-antenna
system can improve the desired channel capacity and impair the
undesired channel capacity simultaneously by making use of its
unique spatial degrees of freedom. Thus, physical layer security
combining with multi-antenna techniques draws considerable research
attentions \cite{Multiantenna2}-\cite{Multiantenna5}. In
\cite{Multiantenna4} and \cite{Multiantenna6}, the optimal
beamforming design methods for MISO and MIMO wiretap channels were
given by maximizing the secrecy rate respectively, assuming that the
multi-antenna information source had full legitimate and
eavesdropper channel state information (CSI). However, in practical
systems, the eavesdropper CSI is difficult to be obtained. It is
proved that if there is no eavesdropper CSI and no artificial noise
in the transmit signals, the beamforming along the direction of the
legitimate channel is optimal \cite{CSIquantization}
\cite{CSIquantization1}. In this case, since the source has no
knowledge of the eavesdropper channel that varies randomly, it is
impossible to maintain a steady secrecy rate over all realization
over fading channels. Hence, the secrecy outage capacity is adopted
as a useful and intuitive metric to evaluate security, which is
defined as the maximum available rate under the condition that the
probability that the real transmission rate is greater than the
secrecy rate is equal to a given value \cite{Outagecapacity}
\cite{Multiantenna7}.

In addition to the achievement of CSI, there are other problems to
be solved in physical layer security. For example, the eavesdropper
may be closer to the information source and has more antennas than
the secure user (SU), such that the secrecy outage capacity is quite
small in order to guarantee the security. In this context, even
though the multi-antenna gain is exploited, it will be a challenging
task to realize a secure, reliable and efficient information
transmission with traditional multi-antenna techniques. Recently,
the concept of large-scale MIMO (LS-MIMO) has been proposed to
significantly improve the performance by taking advantage of its
large array gain \cite{LargescaleMIMO}. Intuitively, due to the high
spatial resolution of LS-MIMO systems, the information leaked to the
eavesdropper is very small. Thus, the information security can be
achieved with a high probability. Inspired by this, we introduce the
LS-MIMO technique into physical layer security. As shown later, by
increasing the number of antennas at the information source,
stringent security requirement can be met even with high QoS
constraint.

However, the use of LS-MIMO is not without challenge. In LS-MIMO
systems, the transmitter commonly obtains the CSI via channel
reciprocity in time division duplex (TDD) systems
\cite{ChannelEstimation}. However, due to a large number of antenna
elements in LS-MIMO systems, channel estimation error is inevitable,
resulting in CSI mismatch and thus performance loss
\cite{ImperfectCSI}. In this paper, we will investigate the effect
of imperfect CSI in LS-MIMO systems on the secrecy outage capacity.

In addition, it is also important for a wireless communications
system to gurantee a certain quality of service (QoS). In
traditional MIMO systems, QoS can always be met by increasing the
transmit power. however, this is not the case in a physical layer
security system. In secure communications based on physical layer
security, the secrecy outage capacity is not an increasing function
of transmit power any more, since both legitimate and eavesdropper
channel capacities improves as transmit power increases
\cite{EnergyEfficientPLS}. Therefore, it makes sense to distribute
the power according to channel conditions, e.g. interception
distance. In this paper, we propose an energy-efficient power
allocation scheme by maximizing the ratio of secrecy outage capacity
and power consumption, namely energy efficiency (bits per Joule),
while fulfilling security and QoS requirements.

The rest of this paper is organized as follows. We first give an
overview of the secure LS-MIMO system and investigate the impact of
imperfect CSI on the secrecy outage capacity in Section II, and then
derive an energy-efficient power allocation scheme by maximizing the
secrecy energy efficiency while satisfying the secrecy, QoS and
power constraints in Section III. In Section IV, we present some
numerical results to validate the effectiveness of the proposed
scheme. Finally, we conclude the whole paper in Section V.

\section{System Model}
\begin{figure}[h] \centering
\includegraphics [width=0.45\textwidth] {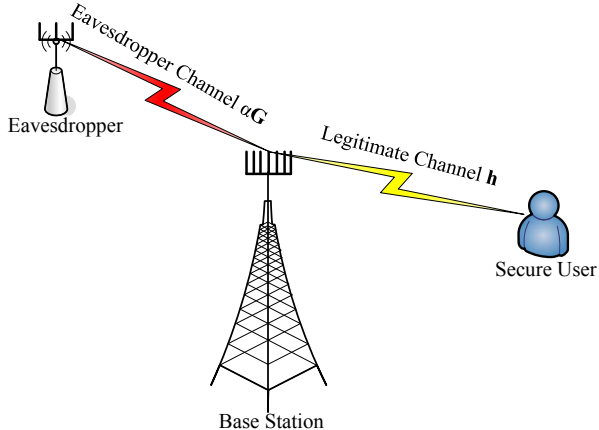}
\caption {An overview of the secure LS-MIMO system model.}
\label{Fig1}
\end{figure}

We consider an LS-MIMO downlink, where a base station (BS) with
$N_t$ antennas communicates with a single antenna secure user (SU),
while a passive eavesdropper with $N_r$ antennas also receives the
signal from the BS due to the broadcast nature of the wireless
channel, and tries to detect the message. Note that the number of BS
antennas $N_t$ can be very large in an LS-MIMO system. We use
$\textbf{h}$ to denote the $N_t$ dimensional legitimate channel
vector from the BS to the SU, where $\textbf{h}$ is a circularly
symmetric complex Gaussian (CSCG) random vector with zero mean and
unit variance. Similarly, we use $\alpha\textbf{G}$ to denote the
$N_r\times N_t$ eavesdropper channel matrix from the BS to the
eavesdropper, where $\alpha$ is the relative path loss and
$\textbf{G}$ is the small scale fading matrix with i.i.d. zero mean
and unit variance complex Gaussian entries, respectively. Note that
a large $\alpha$ means that the eavesdropper is close to the BS, and
thus the interception ability of the eavesdropper is strong. If
$\alpha=1$, the eavesdropper has the same access distance as the SU.
It is assumed that $\alpha$ remains constant during a relatively
long time period due to its slow fading and is known at the BS,
while $\textbf{h}$ and $\textbf{G}$ remain constant in a time slot
and independently fade slot by slot. The network is operated in time
division duplex (TDD) mode, so the channel reciprocity holds true.

At the beginning of each time slot, the SU transmits the pilot for
channel estimation at the BS. Due to a large number of antennas and
limited channel estimation capacity, there may be channel estimation
error. Commonly, the CSI mismatch caused by estimation error can be
modeled as \cite{CEE}
\begin{equation}
\textbf{h}=\sqrt{\rho}\hat{\textbf{h}}+\sqrt{1-\rho}\textbf{e},\label{eqn1}
\end{equation}
where $\hat{\textbf{h}}$ is the estimated CSI with the same
distribution as $\textbf{h}$, and $\textbf{e}$ is the estimation
error noise with i.i.d. zero mean and unit variance complex Gaussian
entries. $\rho$, scaling from 0 to 1, is the correlation coefficient
between $\textbf{h}$ and $\hat{\textbf{h}}$. The larger the
correlation coefficient, the more accurate the estimation. If
$\rho=1$, the BS has perfect CSI. We assume the BS knows the
statistics $\rho$. Note that the BS always has partial CSI by making
use of channel reciprocity, so we do not consider the case of
$\rho=0$ in this paper.

According to the estimated CSI $\hat{\textbf{h}}$, the BS performs
maximum ratio transmission (MRT), namely designing the transmit beam
$\textbf{w}=\frac{\hat{\textbf{h}}}{\|\hat{\textbf{h}}\|}$. Thus,
the received signals at the SU and the eavesdropper are given by
\begin{equation}
y_s=\sqrt{P}\textbf{h}^H\textbf{w}x+n_s\label{eqn2}
\end{equation}
and
\begin{equation}
\textbf{y}_e=\sqrt{P}\alpha\textbf{G}\textbf{w}x+\textbf{n}_e,\label{eqn3}
\end{equation}
respectively, where $x$ is the Gaussian distributed transmit
signal with unit variance, $P$ is the transmit power, $n_s$ and
$\textbf{n}_e$ are the additive Gaussian white noises with unit
variance at the SU and the eavesdropper, respectively. While there are many available methods
to detect the intercepted signal $\textbf{y}_e$ at the eavesdropper, e.g. antenna
selection (AS), maximum ratio combination (MRC), zero-forcing (ZF) and minimum mean square
error (MMSE), we only consider the AS in this paper. Specifically, during each time
slot, the eavesdropper selects one antenna with the strongest gain
to receive and detect the intercepted signal. Hence, the
capacities of the legitimate and eavesdropper channels can be
expressed as
\begin{equation}
C_s=W\log_2(1+\gamma_s)\label{eqn4}
\end{equation}
and
\begin{equation}
C_e=W\log_2(1+\gamma_e),\label{eqn5}
\end{equation}
where $W$ is the spectrum bandwidth,
$\gamma_s=P|\textbf{h}^H\textbf{w}|^2$ and
$\gamma_e=\max\limits_{1\leq i\leq
N_r}P\alpha^2|\textbf{g}_i\textbf{w}|^2$ are the signal-to-noise
ratio (SNR) at the SU and the eavesdropper respectively, and
$\textbf{g}_i$ is the $i$th row of $\textbf{G}$. For the legitimate
channel capacity $C_s$ in an LS-MIMO system with imperfect CSI, we
have the following lemma:

\emph{Lemma 1}: As $N_t$ approaches infinity, the legitimate channel
capacity $C_s$ asymptotically approaches $W\log_2(\rho PN_t)$.

\begin{proof}
Please refer to the Appendix.
\end{proof}

From an information-theoretic view, the secrecy rate is given by
$C_{sec}=[C_s-C_e]^{+}$, where $[x]^{+}=\max(x,0)$. Since there is
no knowledge of the eavesdropper channel at the BS, it is impossible
to maintain a steady secrecy capacity over all realizations over
fading channels. In this paper, we take the secrecy outage capacity
$R_{sec}$ as the performance metric, which is defined as the maximum
achievable rate under the condition that the outage probability that
the transmission rate surpasses the secrecy rate is equal to a given
value $\varepsilon$, namely
\begin{equation}
P_r\left(R_{sec}>C_s-C_e\right)=\varepsilon.\label{eqn7}
\end{equation}
Substituting (\ref{eqn5}) and (\ref{app4}) into (\ref{eqn7}), the
outage probability can be transformed as
\begin{eqnarray}
\varepsilon&=&P_r\left(\gamma_e>\rho PN_t2^{-R_{sec}/W}-1\right)\nonumber\\
&=&1-F_{\gamma_{e}}\left(\rho
PN_t2^{-R_{sec}/W}-1\right),\label{eqn8}
\end{eqnarray}
where $F_{\gamma_{e}}(x)$ is the cumulative distribution function
(cdf) of $\gamma_e$. Since $\textbf{w}$ is independent of
$\textbf{g}_j$ for an arbitrary $j$, $|\textbf{g}_j\textbf{w}|^2$
is exponentially distributed. Thus, according to the order theory,
we have
\begin{equation}
F_{\gamma_{e}}(x)=\left(1-\exp\left(-\frac{x}{P\alpha^2}\right)\right)^{N_r}.\label{eqn9}
\end{equation}
Substituting (\ref{eqn9}) into (\ref{eqn8}), it is obtained that
\begin{eqnarray}
\varepsilon=1-\left(1-\exp\left(-\frac{\rho
PN_t2^{-R_{sec}/W}-1}{P\alpha^2}\right)\right)^{N_r}.\label{eqn10}
\end{eqnarray}

\emph{Remark}: The secrecy outage probability $\varepsilon$ is a
monotonically increasing function of $N_r$ and a monotonically
decreasing function of $N_t$. As $N_r$ increases, the secrecy outage
probability $\varepsilon$ will increase for a given secrecy outage
capacity $R_{sec}$. To solve this problem, by letting
$N_t\rightarrow\infty$, $\varepsilon$ asymptotically approaches 0,
the security requirement can be guaranteed nearly with probability
1. Hence, in an LS-MIMO system, when the BS is equipped with a large
number of antennas, even the eavesdropper is closer to the BS (i.e.
$\alpha>1$) and has more antennas than the SU, the BS can still
support a high transmission rate with a low outage probability, and
then meets both secrecy and QoS requirements.

Given transmit power $P$ and the requirement of outage probability
$\varepsilon$, the secrecy outage capacity can be expressed as
\begin{equation}
R_{sec}=-W\log_2\left(\frac{1-P\alpha^2\ln\left(1-(1-\varepsilon)^{1/N_r}\right)}{\rho
PN_t}\right).\label{eqn11}
\end{equation}
From (\ref{eqn10}), we can also obtain the probability of positive
secrecy capacity as
\begin{eqnarray}
P_r(C_{sec}>0)=\left(1-\exp\left(-\frac{\rho
PN_t-1}{P\alpha^2}\right)\right)^{N_r}.\label{eqn12}
\end{eqnarray}

Intuitively, the probability $P_r(C_{sec}>0)$ increases as $N_t$
increases and decreases as $\alpha$ and $N_r$ increase, which also
proves that we can easily solve the problems that the eavesdropper
is closer to the BS and has more antennas than the SU by adding the
antennas at the BS in an LS-MIMO system.

\section{Energy-Efficient Power Allocation}
Considering the demand for green communication, we attempt to derive
a secure and efficient power allocation scheme to maximize the
secrecy energy efficiency while satisfying the secrecy, QoS and
power constraints, which is equivalent to the following optimization
problem:
\begin{eqnarray}
J_1:\max &&\frac{R_{sec}}{P_0+P}\label{eqn14}\\
\textmd{s.t.}&& \varepsilon\leq\varepsilon_{\max}\label{eqn15}\\
&&R_{sec}\geq R_{\min}\label{eqn16}\\
&&P\leq P_{\max},\label{eqn17}
\end{eqnarray}
where $P$ is the transmit power as defined in Section II, and $P_0$
is the constant power consumption in the transmit filter, mixer,
frequency synthesizer and digital-to-analog converter, which are
independent of the actual transmit power. (\ref{eqn14}) is the so
called secrecy energy efficiency, defined as the number of
transmission bits per Joule. (\ref{eqn15}) is used to fulfill the
secrecy requirement based on physical layer security, and
(\ref{eqn16}) is the QoS constraint, where $R_{\min}$ is the minimum
effective transmission rate to meet a given QoS requirement, such as
delay provisioning. $P_{\max}$ is the constraint on maximum transmit
power. Since $\varepsilon$ is a monotonically increasing function of
$R_{sec}$ and a decreasing function of $P$, the condition of
$\varepsilon=\varepsilon_{\max}$ is optimal in the sense of
maximizing the secrecy energy efficiency. Thus, (\ref{eqn15}) can be
canceled and $R_{sec}$ can be replaced by
$-W\log_2\left(\frac{1-P\alpha^2\ln\left(1-(1-\varepsilon_{\max})^{1/N_r}\right)}{\rho
PN_t}\right)$. Notice that there may be no feasible solution for
$J_1$, due to the stringent secrecy, QoS and power constraints.
Under such a condition, in order to obtain a solution, that will
fulfill the requirements, we can increase the number of antennas at
the BS, which is the promising advantage of the LS-MIMO system.

The objective function (\ref{eqn14}) in a fractional program is a
ratio of two functions of the optimization variable $P$, resulting
in that $J_1$ is a fractional programming problem, which is in
general nonconvex. Following \cite{EnergyEfficientPLS}, the
objective function is equivalent to
$-W\log_2\left(\frac{1-P\alpha^2\ln\left(1-(1-\varepsilon_{\max})^{1/N_r}\right)}{\rho
PN_t}\right)-q^{\star}(P_0+P)$ by exploiting the properties of
fractional programming, where $q^{\star}$ is the secrecy energy
efficiency when $P$ is equal to the optimal power $P^{\star}$ of
$J_1$, namely
$q^{\star}=\frac{-W\log_2\left(\frac{1-P^{\star}\alpha^2\ln\left(1-(1-\varepsilon_{\max})^{1/N_r}\right)}{\rho
P^{\star}N_t}\right)}{(P_0+P^{\star})}$. Thus, $J_1$ is transformed
as
\begin{eqnarray}
J_2\!\!\!\!\!&:&\!\!\! \min
W\log_2\left(\frac{1-P\alpha^2\ln\left(1-(1-\varepsilon_{\max})^{1/N_r}\right)}{\rho
PN_t}\right)\nonumber\\
&&\quad\quad+q^{\star}(P_0+P)\label{eqn18}\\
\textmd{s.t.}&& P\geq\frac{1}{\alpha^2\ln\left(1-(1-\varepsilon_{\max})^{1/N_r}\right)+\rho N_t2^{-R_{\min}/W}}\nonumber\\
&&\quad=P_{\min}\label{eqn19}\\
&&P\leq P_{\max}.\label{eqn20}
\end{eqnarray}
$J_2$, as a convex optimization problem, can be solved by the
Lagrange multiplier method. By some arrangement, its Lagrange dual
function can be written as
\begin{eqnarray}
\mathcal{L}(\mu,\nu,P)&=&W\log_2\left(\frac{1-P\alpha^2\ln\left(1-(1-\varepsilon_{\max})^{1/N_r}\right)}{\rho
PN_t}\right)\nonumber\\&+&q^{\star}(P_0+P)-\mu P+\mu P_{\min}+\nu
P-\nu P_{\max},\nonumber\\\label{eqn21}
\end{eqnarray}
where $\mu\geq0$ and $\nu\geq0$ are the Lagrange multipliers
corresponding to the constraint (\ref{eqn19}) and (\ref{eqn20}),
respectively. Therefore, the dual problem of $J_2$ is given by
\begin{eqnarray}
\max\limits_{\mu,\nu}\min\limits_{P}\mathcal{L}(\mu,\nu,P).\label{eqn22}
\end{eqnarray}
Given $\mu$ and $\nu$, the optimal power $P^{\star}$ can be derived
by solving the following KKT condition
\begin{eqnarray}
\frac{\partial\mathcal{L}(\mu,\nu,P)}{\partial
P}&=&\frac{W}{\ln2(P^2\alpha^2\ln\left(1-(1-\varepsilon_{\max})^{1/N_r}\right)-P)}\nonumber\\
&&-q^{\star}-\mu+\nu=0.\label{eqn23}
\end{eqnarray}
Moreover, $\mu$ and $\nu$ can be updated by the gradient method,
which are given by
\begin{equation}
\mu(n+1)=[\mu(n)-\triangle_{\mu}(P_{\min}-P)]^{+}\label{eqn24}
\end{equation}
and
\begin{equation}
\nu(n+1)=[\nu(n)-\triangle_{\nu}(P-P_{\max})]^{+},\label{eqn25}
\end{equation}
where $n$ is the iteration index, and $\triangle_{\mu}$ and
$\triangle_{\nu}$ are the positive iteration steps. Inspired by the
Dinkelbach method \cite{Dinkelbach}, we propose an iterative
algorithm as follows\\
\rule{8.82cm}{1pt}\\
Algorithm 1: Energy-Efficient Power Allocation\\
\rule{8.82cm}{1pt}
\begin{enumerate}

\item Initialization: Given $N_t$, $N_r$ $W$, $\alpha$, $\rho$,
$R_{\min}$, $P_0$, $P_{\max}$, $\triangle_{\mu}$,
$\triangle_{\nu}$ and $\varepsilon_{\max}$. Let $\mu=0$, $\nu=0$,
$P=0$ and
$q^{\star}=-W\log_2\left(\frac{1-P\alpha^2\ln\left(1-(1-\varepsilon_{\max})^{1/N_r}\right)}{\rho
PN_t}\right)/(P_0+P)$. $\epsilon$ is a sufficiently small positive
real number.

\item Update $\mu$ and $\nu$ according to (\ref{eqn24}) and
(\ref{eqn25}), respectively.

\item Computing the optimal $P^{\star}$ by solving the equation (\ref{eqn23}).

\item If
$-W\log_2\left(\frac{1-P^{\star}\alpha^2\ln\left(1-(1-\varepsilon_{\max})^{1/N_r}\right)}{\rho
P^{\star}N_t}\right)-q^{\star}(P_0+P^{\star})>\epsilon$, then set
$q^{\star}=\\
-W\log_2\left(\frac{1-P^{\star}\alpha^2\ln\left(1-(1-\varepsilon_{\max})^{1/N_r}\right)}{\rho
P^{\star}N_t}\right)/(P_0+P^{\star})$, and go to 2). Otherwise,
$P^{\star}$ is the optimal transmit power.
\end{enumerate}
\rule{8.82cm}{1pt}

\section{Numerical Results}
To examine the effectiveness of the proposed energy-efficient power
allocation scheme, we present several numerical results in the
following scenarios: we set $N_t=20$, $N_r=2$, $W=1$MHz,
$R_{\min}=1.5$Mb/s, $P_0=0.5$Watt and $P_{\max}=10$Watt. It is found
that the proposed energy-efficient power allocation scheme converges
after no more than 20 times iterative computation in all simulation
scenarios.

\begin{figure}[h] \centering
\includegraphics [width=0.5\textwidth] {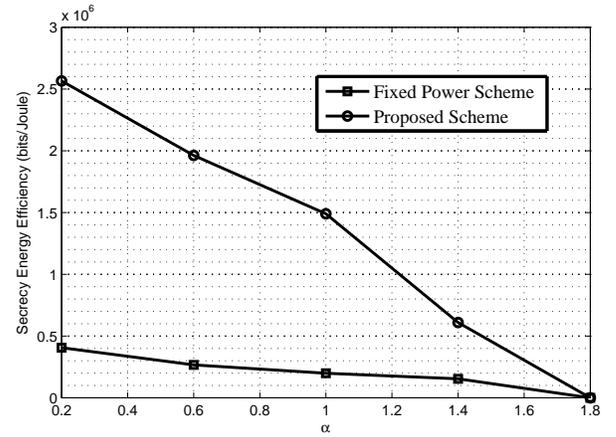}
\caption {Performance comparison of the fixed and the proposed power
allocation schemes.} \label{Fig2}
\end{figure}

Fig.\ref{Fig2} compares the secrecy energy efficiency of the
proposed energy-efficient and the fixed power allocation schemes
with $\varepsilon_{\max}=0.05$, $\rho=0.8$ and $N_t=20$. Note that
we set $P=P_{\max}$ fixedly for the fixed scheme. As seen in
Fig.\ref{Fig2}, the proposed scheme performs better than the fixed
one, especially when $\alpha$ is small. For example, when
$\alpha=1$, there is about $1.3$Mb/J gain. It is worth pointing
out that our proposed scheme uses less power than the fixed
scheme. Therefore, the proposed scheme is more suitable for the
future green and secure communications.

\begin{figure}[h] \centering
\includegraphics [width=0.5\textwidth] {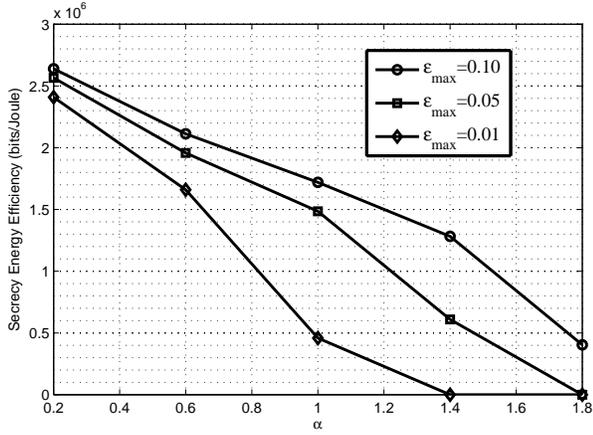}
\caption {Performance comparison of the proposed power allocation
scheme with different requirements of secrecy outage probability.}
\label{Fig3}
\end{figure}

\begin{figure}[h] \centering
\includegraphics [width=0.5\textwidth] {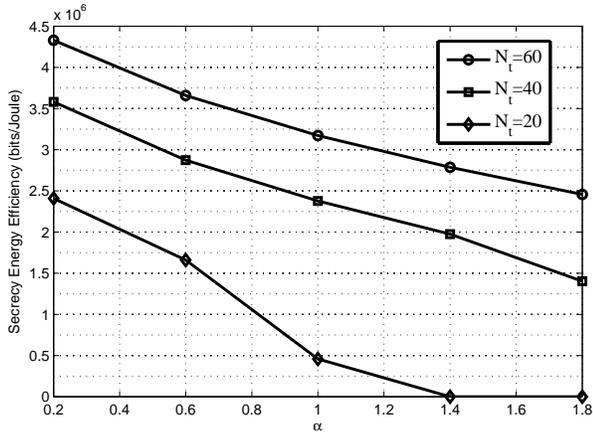}
\caption {Performance comparison of the proposed power allocation
scheme with different number of antennas at the BS.} \label{Fig4}
\end{figure}

Fig.\ref{Fig3} investigates the effect of the outage probability
requirements on the secrecy energy efficiency of the proposed
scheme with $\rho=0.8$ and $N_t=20$. For a given $\alpha$, as
$\varepsilon_{\max}$ decreases, the secrecy energy efficiency
reduces accordingly, this is because more power is used to
decrease the outage probability. On the other hand, for a given
outage probability requirement, the increase of $\alpha$ leads to
the decrease of the secrecy energy efficiency, since the
eavesdropper has a strong eavesdropping ability. It is found that
when $\alpha=1.4$ and $\varepsilon_{\max}=0.01$, the secrecy
energy efficiency reduces to zero. This is because there is no
nonzero secrecy outage capacity under such conditions. Such a
challenging problem can be easily solved in LS-MIMO systems by
adding the antennas at the BS. As seen in Fig.\ref{Fig4}, with the
increase of the number of antennas, the secrecy energy efficiency
increases significantly, especially when $\alpha$ is large. Thus,
a secure, reliable and QoS guaranteed communication can be
realized even with short-distance interception.

\begin{figure}[h] \centering
\includegraphics [width=0.5\textwidth] {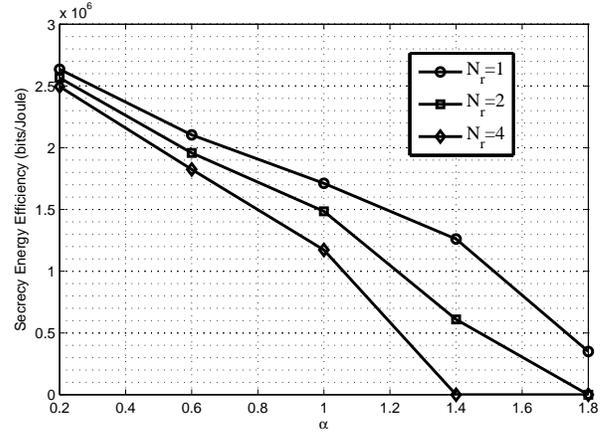}
\caption {Performance comparison of the proposed power allocation
scheme with different number of antennas at the eavesdropper.}
\label{Fig6}
\end{figure}

Then, we show the impact of the number of antennas at the
eavesdropper on the secrecy energy efficiency of the proposed
scheme with $\varepsilon_{\max}=0.05$, $\rho=0.8$ and $N_t=20$. As
seen in Fig.\ref{Fig6}, with the increase of $N_r$, the energy
efficiency decreases according, especially in the region of large
$\alpha$. At $\alpha=1.4$, the secrecy energy efficiency with
$N_r=4$ reduces to zero. Fortunately, this problem can be solved
by adding the number of antennas at the BS. Hence, the proposed
scheme can address the challenge that the eavesdropper has more
antennas than the SU.

\begin{figure}[h] \centering
\includegraphics [width=0.5\textwidth] {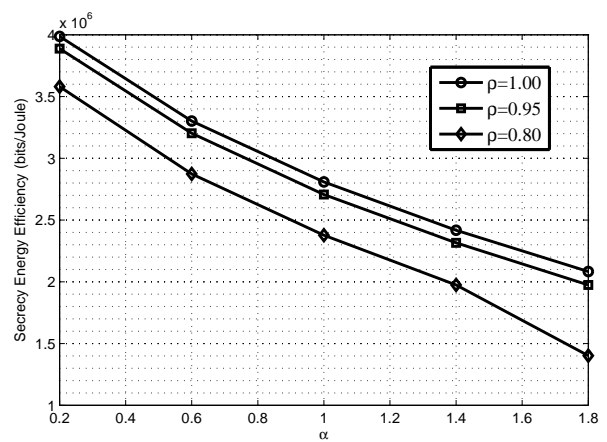}
\caption {Performance comparison of the proposed power allocation
scheme with different correlation coefficients.} \label{Fig5}
\end{figure}

Finally, we investigate the impact of imperfect CSI on the secrecy
energy efficiency of the proposed scheme with
$\varepsilon_{\max}=0.01$ and $N_t=40$. As shown in Fig.\ref{Fig5},
even with slight CSI mismatch, i.e. $\rho=0.95$, there is
performance loss with respect to the ideal case of $\rho=1$. With
the increase of $\rho$, secrecy energy efficiency decreases
accordingly.

\section{Conclusion}
In this paper, we show that the large number of antenna elements
of LS-MIMO system can address the problem of guaranteeing both
stringent security and high QoS requirements when the eavesdropper
is closer to the BS and has more antennas than the SU. The impact
of imperfect CSI at the BS is investigated quantitatively. We also
design an energy-efficient power allocation scheme to enable a
secure communication while satisfy the QoS constraint. Our
analysis and numerical results show that LS-MIMO is an
energy-efficient technique that provides physical layer security
even under some adverse conditions.

\appendix
By using the transmit beam for MRT
$\textbf{w}=\hat{\textbf{h}}/\|\hat{\textbf{h}}\|$, the legitimate
channel capacity can be expressed as
\begin{eqnarray}
C_s&=&W\log_2\left(1+P\left|\left(\sqrt{\rho}\hat{\textbf{h}}+\sqrt{1-\rho}\textbf{e}\right)^H\frac{\hat{\textbf{h}}}{\|\hat{\textbf{h}}\|}\right|^2\right)\nonumber
\end{eqnarray}
\begin{eqnarray}
&=&W\log\bigg(1+\rho
P\|\hat{\textbf{h}}\|^2+2\rho(1-\rho)P\mathcal{R}(\textbf{e}^H\hat{\textbf{h}})\nonumber\\
&&+(1-\rho)P\|\textbf{e}\hat{\textbf{h}}^H\|^2/\|\hat{\textbf{h}}\|^2\bigg)\label{app1}\\
&\approx&W\log_2(1+\rho P\|\hat{\textbf{h}}\|^2)\label{app2}\\
&\approx&W\log_2(1+\rho PN_t)\label{app3}\\
&\approx&W\log_2(\rho PN_t),\label{app4}
\end{eqnarray}
where $\mathcal{R}(x)$ denotes the real part of $x$. (\ref{app2})
follows the fact that $\rho P\|\hat{\textbf{h}}\|^2$ scales with the
order $\mathcal{O}(\rho PN_t)$ as $N_t\rightarrow\infty$ while
$2\rho(1-\rho)P\mathcal{R}(\textbf{e}^H\hat{\textbf{h}})+
\frac{(1-\rho)P\|\textbf{e}\hat{\textbf{h}}^H\|^2}{\|\hat{\textbf{h}}\|^2}$
scales as the order $\mathcal{O}(1)$, so the third and forth terms
can be negligible. (\ref{app3}) holds true because of
$\lim\limits_{N_t\rightarrow\infty}\frac{\|\hat{\textbf{h}}\|^2}{N_t}=1$,
and (\ref{app4}) neglects the constant 1 in the case of large $N_t$.
Therefore, we get the Lemma 1.

\end{document}